\documentclass[11pt]{article}
\usepackage{moriond,amssymb,hyperref}
\usepackage[dvips]{graphicx}

\def\chibarchi{\left\langle\overline{\chi}\chi\right\rangle}

\begin{document}

\vspace*{4cm}

\title{LATTICE SUPERSYMMETRY WITH DOMAIN WALL FERMIONS}

\author{GEORGE T.\ FLEMING}

\address{Physics Department, The Ohio State University,\\
         Columbus OH 43210-1168, USA}

\maketitle\abstracts{Supersymmetry, like Poincar\'{e} symmetry, is
softly broken at finite lattice spacing provided the gaugino mass term is
strongly suppressed.  Domain wall fermions provide the mechanism
for suppressing this term by approximately imposing chiral symmetry
on the lattice.  We present the first numerical simulations of ${\cal N}=1$
supersymmetric ${\rm SU}(2)$ Yang-Mills on the lattice in $d=4$ dimensions
using domain wall fermions.}

\section{Introduction}
\label{sec:Introduction}

Supersymmetric (SUSY) field theories may play an important role
in describing the physics beyond the Standard Model.  Non-perturbative
numerical studies of these theories could provide confirmation
of existing analytical calculations and new insights on aspects of the theories
not currently accessible to analytic methods, similar in spirit to the current
situation in many other field theories, most notably QCD.  Several,
but not all, SUSY theories can be formulated and studied numerically
on the lattice \cite{Curci:1987sm,Neuberger:1998bg,Kaplan:2000jn}.
Foremost, as the lattice spacetime is discreet, only a discrete subgroup
of the Poincar\'{e} symmetry is unbroken, so any formulation will break SUSY.
However, the allowed operators that break Poincar\'{e} symmetry are irrelevant
in the continuum limit, so one can calculate at several lattice spacings $a$
and take the $a\to 0$ limit.

If a SUSY model has non-trivial scalar fields, scalar mass terms will break
SUSY unless forbidden by some symmetry.  Since these operators are relevant
fine tuning will be needed to cancel their contributions
in the continuum limit.  We avoid this problem as the four-dimensional
${\cal N}=1$ Super Yang-Mills (SYM) theory does not involve scalars.
The lattice fermion doubling problem will also break SUSY due to a mismatch
in the number of bosonic and fermionic degrees of freedom.  Removing doublers
breaks chiral symmetry, allowing relevant gluino mass terms.
In the traditional approach \cite{Curci:1987sm}, fine tuning is used
to cancel these mass terms.  Pioneering work using these methods has
already produced very interesting numerical results
\cite{Campos:1999du,Donini:1998hh}.

Many properties of the ${\cal N}=1$ SYM have already been computed
using analytic techniques that take particular advantage of the supersymmetry
in the model.  On the contrary, numerical simulations of SYM are at least
as computationally difficult as dynamical QCD.  So, one must be
careful to choose an interesting problem since the trade-off is to not
study some aspect of dynamical QCD.  For us, one such interesting problem is
to determine if the vacuum supports a non-zero gluino condensate $\chibarchi$,
as widely believed, and whether the formation of the condensate is due
to spontaneous or anomalous symmetry breaking.

The Dirac operator in the adjoint representation of ${\rm SU}(N)$ has an index
$2N\nu$, where $\nu$ is the winding of the gauge field.  Classical instantons
have integer winding and cause condensations of operators with $2N$ gluinos
which anomalously breaks the ${\rm U}(1)$ $R$-symmetry to ${\rm Z}_{2N}$.
For $\chibarchi$ to condense, the remaining ${\rm Z}_{2N}$ symmetry must
further break either spontaneously or anomalously to $Z_2$.  If the breaking
is anomalous, then the responsible gauge configurations must have fractional
winding \cite{Fleming:2000fa}.  It has already been established that such
gauge configurations do exist \cite{Edwards:1998dj}.  It is our goal
to distinguish between these two scenarios.

This work summarizes results recently presented by the author in collaboration
with John B.\ Kogut and Pavlos M.\ Vranas \cite{Fleming:2000fa}.  All numerical
simulations were run on QCDSP supercomputers at Columbia Univ.\
and Ohio State.  For reviews on DWF please see the LATTICE '00 review talk
of Vranas \cite{Vranas:2000} and references therein.  The possible use of DWF
in SUSY theories has been discussed in earlier works
\cite{Neuberger:1998bg,Kaplan:2000jn} and the methods used here
are along these lines.  For lists of references not included here for lack
of space, please see the cited articles \cite{Fleming:2000fa}.

\section{The DWF formulation}
\label{sec:The_DWF_formulation}

In the lattice DWF formulation of a vector-like theory in $d=4$ dimensions,
the fermionic fields are defined on a $d+1$ dimensional lattice using
a local action.  It is often convenient to treat the extra dimension
as a new internal flavor space.  Thus, the gauge fields are introduced
in the standard way in the $d$ dimensional spacetime and are coupled
to the extra fermion degrees of freedom like extra flavors.  The detailed form
of the action can be found in our recent work \cite{Fleming:2000fa}.

The key ingredient is the free boundary conditions imposed on the extra
dimension.  As a result, two chiral exponentially bound surface states
appear on the boundaries (domain walls) with the plus chirality localized
on one wall and the minus chirality on the other.  The two chiralities mix
only by an amount that is exponentially small in the size
of the extra dimension, called $L_s$, and together form a Dirac spinor
that propagates in the $d$ dimensional spacetime with an exponentially small
mass.  Therefore, the chiral symmetry breaking artificially induced
by the Wilson term can be controlled by the new parameter $L_s$.
It is often convenient to introduce an additional bare gluino mass $m_f$
to help control extrapolations to the chiral limit.
In the $L_s\to\infty, m_f\to 0$ limit chiral symmetry is exact at any lattice
spacing without fine tuning.

The computing requirement is linear in $L_s$, in contrast to traditional
lattice fermion regulators where the chiral limit is approached
only as the continuum limit is taken, a process that is achieved
at a large computing cost.  Specifically, because of algorithmic reasons,
the computing cost to reduce the lattice spacing by a factor of two grows
by a factor of $2^{8-10}$ in four dimensions.  Therefore, the unique properties
of DWF provide a way to bring under control the systematic chiral symmetry
breaking effects using today's supercomputers.

The application of DWF to ${\cal N}=1\ {\rm SU}(N)$ SYM is quite similar
to $N_f=1\ {\rm SU}(N)$ QCD.  The differences are merely that the fermions are
in the adjoint representation of ${\rm SU}(N)$ and that the Dirac fermion fields
must satisfy the Majorana constraint, thereby reducing by half the number
of fermion degrees of freedom.  After integrating out the Grassmann fields
subject to the Majorana constraint, the Dirac determinant is replaced
by a Pfaffian:  essentially the square root of the determinant,
provided it is positive.  For DWF, the Pfaffian cannot change sign so working
with either Pfaffians or square roots of determinants is equivalent.

\section{Numerical results}
\label{sec:numerical_results}

All numerical simulations were performed on $8^4$ and $4^4$ spacetime
volumes using the inexact hybrid molecular dynamics (HMD) $R$ algorithm.
The algorithm numerically integrates the classical equations of motion
as part of generating a statistical ensemble with weights proportional
to the fourth root of a two adjoint flavor Dirac determinant.
For DWF, this weight is proportional to a single adjoint Majorana flavor.
The step sizes were chosen such that systematic uncertainties
due to numerical integration errors are negligible compared
to statistical uncertainties.

\begin{figure}[h]
\hfill
\begin{minipage}{0.4\textwidth}
  \vspace{-1.0cm}
  \includegraphics[width=\textwidth]{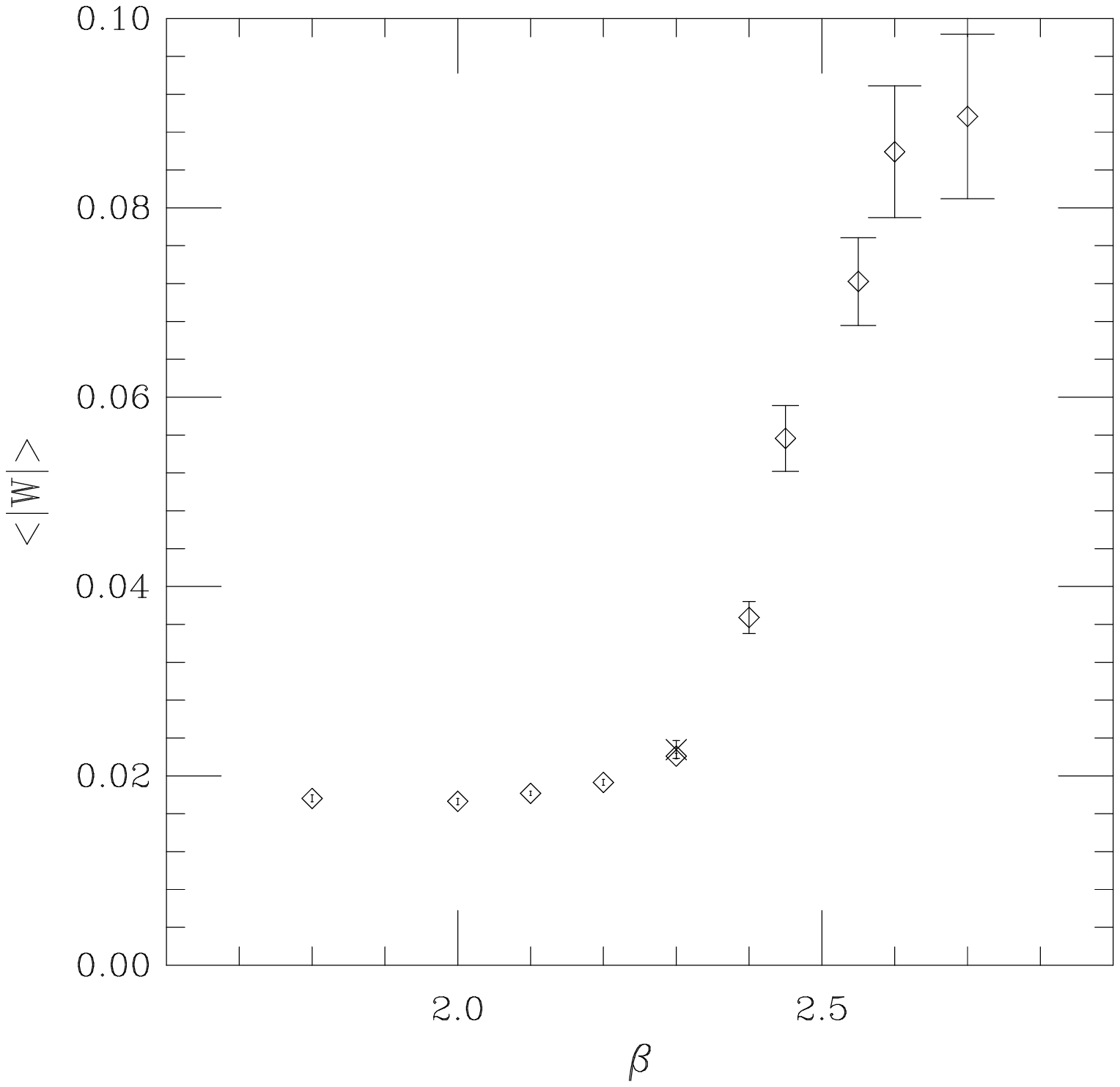}
  \vspace{-1.0cm}
  \caption{
    Wilson line $\left\langle\left|W\right|\right\rangle$ on an $8^4$ lattice.
    Diamonds are quenched and the cross is dynamical with $L_s=24$ and $m_f=0$.
  }
  \label{fig:one}
\end{minipage}
\qquad
\begin{minipage}{0.4\textwidth}
  \includegraphics[width=\textwidth]{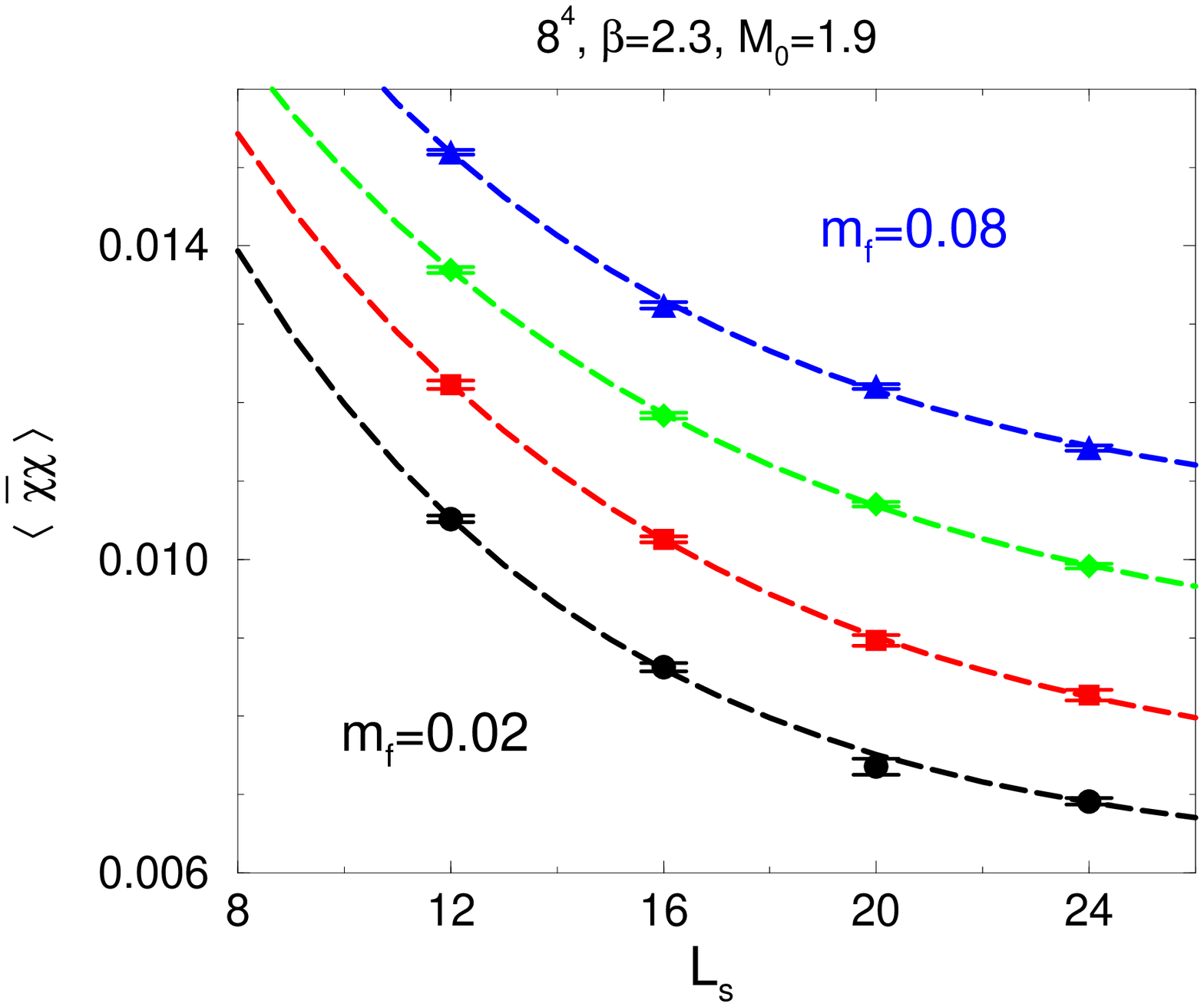}
  \vspace{-1.0cm}
  \caption{
    Gluino condensate $\chibarchi$ {\it vs}.\ $L_s$ with $\beta=2.3$ and
    $m_f=0.02,0.04,0.06,0.08$.  Curves are an exponential fit.
  }
  \label{fig:two}
\end{minipage}
\hfill\mbox{}
\end{figure}

The $8^4$ volume simulations were done with $\beta\equiv 4/g^2=2.3$.  The value
of $\beta$ was chosen as large as possible without entering the thermal
transition region.  From quenched simulations in an $8^4$ volume, the magnitude
of the fundamental Wilson line $\left\langle\left|W\right|\right\rangle$,
the quenched order parameter, is plotted {\it vs}.\ $\beta$
in figure \ref{fig:one}.  The value
of $\left\langle\left|W\right|\right\rangle$ from a simulation
of the dynamical theory at $\beta=2.3$ is also shown (cross),
indicating that the dynamical theory is in a phase that ``confines''
fundamental sources.  Therefore, the box size is large enough to avoid
thermal effects that break SUSY.  Using similar arguments, the $4^4$ volume
simulations were done with $\beta=2.1$, near the limit of the weak coupling
regime.  Scaling arguments appropriate for weak coupling suggest that
the lattice spacing at $\beta=2.1$ is twice as large as at $\beta=2.3$.

To extrapolate the measured values of $\chibarchi$
to the chiral limit, $L_s\to\infty$ and $m_f\to 0$, simulations were performed
in $8^4$ volumes at fixed $\beta=2.3$ while the size
of the extra dimension $L_s$ was varied between 12 and 24 and the bare mass
$m_f$ was varied between 0.02 and 0.08.  The measured values appear as
the points in figure \ref{fig:two}.  According to Leutwyler and Smilga
\cite{Leutwyler:1992yt}, if the formation of the gluino condensate
is due to spontaneous symmetry breaking, the lattice volume limits how small
the dynamical qluino mass can be set without losing the condensate:
$12 m_{\rm eff} \chibarchi V \gg 1$ (the 12 is just normalization).
As $m_{\rm eff} \gtrsim m_f$, this limit is satisfied for all $8^4$
simulations with $m_f\ge 0.02$.

\begin{figure}[ht]
\hfill
\begin{minipage}{0.4\textwidth}
  \includegraphics[width=\textwidth]{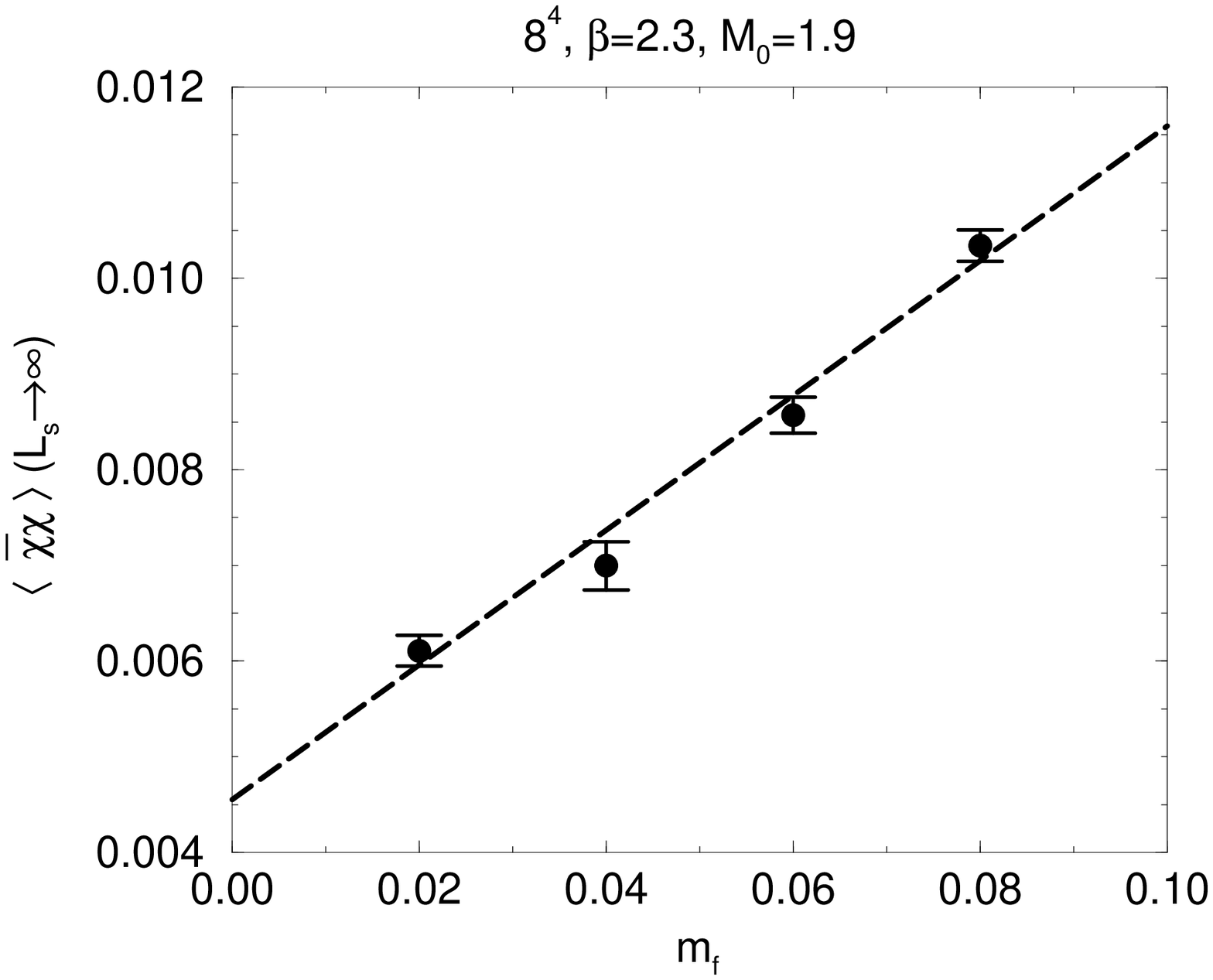}
  \vspace{-1.0cm}
  \caption{
    Extrapolated $\chibarchi$ to $L_s\to\infty$ limit {\it vs}.\ $m_f$
    and linear fit to $m_f\to 0$ limit.
  }
  \label{fig:three}
\end{minipage}
\qquad
\begin{minipage}{0.4\textwidth}
  \includegraphics[width=\textwidth]{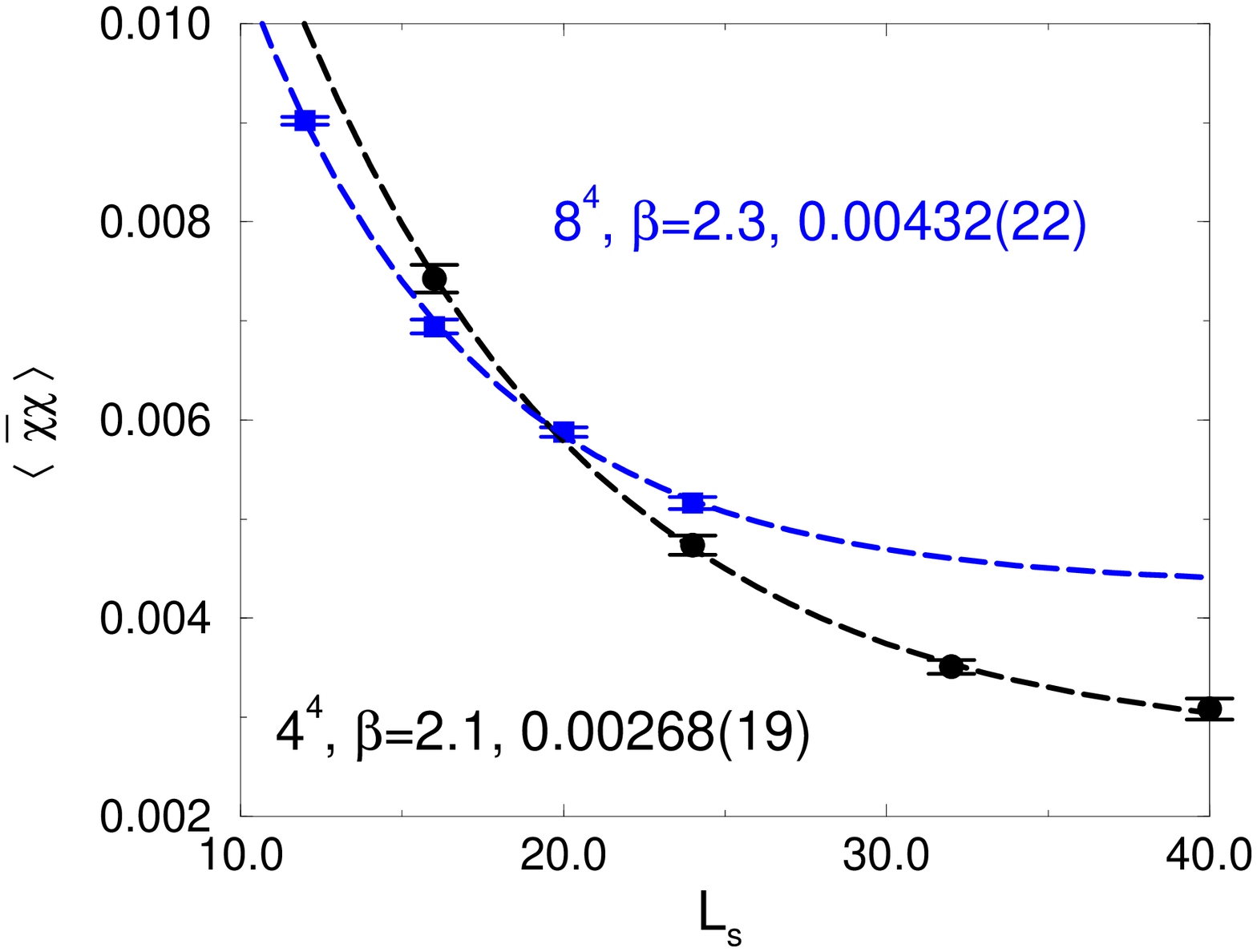}
  \vspace{-1.0cm}
  \caption{
    Dynamical $\chibarchi$ at $m_f=0$ {\it vs}.\ $L_s$ on an $8^4$ lattice
    at $\beta=2.3$ and a $4^4$ lattice with $\beta=2.1$.
    Curves are exponential fits.
  }
  \label{fig:four}
\end{minipage}
\hfill\mbox{}
\end{figure}

To estimate the gluino condensate in the chiral limit, we first extrapolate
at fixed $m_f$ to the $L_s\to\infty$ limit using the fit function
$\chibarchi = c_0 + c_1 \exp(-c_2 L_s)$.
The best fit functions are plotted as the curves in figure \ref{fig:two}.
The values of the extrapolated gluino condensate (with propagated errors)
appear as points in figure \ref{fig:three}.  These extrapolated values
are then further extraploted to the $m_f\to 0$ limit using a linear function
$\chibarchi\left(L_s\to\infty\right) = b_0 + b_1 m_f$.
The best fit function appears as the line in figure \ref{fig:three}.
It is also reassuring to note that reversing the order of limits,
{\it i.~e.\ } first $m_f\to 0$ at fixed $L_s$ then $L_s\to\infty$, yields
a statistically consistent answer.

Another approach to estimating the gluino condensate in the chiral limit
is to actually perform dynamical simulations with $m_f=0$.
Since finite $L_s$ will induce an exponentially small breaking of chiral
symmetry, the effective gluino mass will not be zero.  However, the gluino
mass should be too small to support spontaneous symmetry breaking.
Additional simulations were run for $L_s=12,16,20,24$.  The data are shown
in figure \ref{fig:four}.  The curve is the best fit to the exponential
fitting function with the extrapolated value of the condensate as shown.

Surprisingly, both methods for estimating the gluino condensate produce
consistent results within the statistical errors. Note that this is
inconsistent with the notion of spontaneous symmetry breaking.  Operationally,
this result reinforces our claim that systematic uncertainties
are still relatively small despite limited statistical precision.  Further,
it gives us some confidence that our fit functions are valid over the region
of interest.

To further check for spontaneous symmetry breaking of the ${\rm Z}_4$ symmetry,
we measured $\chibarchi$ on smaller $4^4$ lattices with $m_f=0$ and
even larger values for $L_s$.  The data are shown in figure \ref{fig:four}
with the best exponential fit and the extrapolated value for the condensate.
This provides even stronger evidence that spontaneous symmetry
breaking is not responsible for the formation of a gluino condensate,
at least in finite volumes.  On these lattices $12 m_f V \chibarchi < 1$,
so analytical considerations \cite{Leutwyler:1992yt} suggest the support
of $\chibarchi$ must come primarily from topological sectors
with fractional winding of $\nu=\pm 1/2$.

The spectrum of the theory is of great interest but it was not possible
to measure it on the small lattices considered here.  Also,
the gluino condensate was measured at only two different lattice spacings
so extrapolation to the continuum limit to compare with analytical results
is not possible.  Future work could explore these very interesting topics.


\end{document}